\documentclass[10pt,a4paper,twocolumn,english,pra,aps,showpacs,floatfix,groupedaddress,superscriptaddress]{revtex4-1}

\usepackage{graphicx}
\usepackage{epsfig}
\usepackage[english]{babel}
\usepackage{amsmath}
\usepackage{amssymb}
\usepackage{amsfonts}
\usepackage{longtable}
\usepackage{dcolumn}
\usepackage{bm}
\usepackage{bbm}
\usepackage{nicefrac}
\usepackage{color,array}
\usepackage{colortbl}

\usepackage[normalem]{ulem}

\newcommand{\ket}[1]{\left|#1\right\rangle}		

\def\tcr#1{\textcolor{black}{#1}}
\def\tcb#1{\textcolor{black}{#1}}

\begin{document}
\title{Time-dependent correlations in quantum magnets at finite temperature}

\author{B.\ Fauseweh}
\email{benedikt.fauseweh@tu-dortmund.de}
\affiliation{Lehrstuhl f\"{u}r Theoretische Physik I, Technische Universit\"at Dortmund, Otto-Hahn Stra\ss{}e 4, 44221 Dortmund, Germany}

\author{F.\ Groitl}
\email{Felix.Groitl@psi.ch}
\affiliation{Laboratory for Quantum Magnetism, \'Ecole Polytechnique F\'ed\'erale de Lausanne, 1015 Lausanne, Switzerland}
\affiliation{Laboratory for Neutron Scattering and Imaging, Paul Scherrer Institute, 5232 Villigen-PSI, Switzerland}

\author{T.\ Keller}
\affiliation{Max Planck Institute For Solid State Research, 70569 Stuttgart, Germany}
\affiliation{Max Planck Society Outstation at the FRM II, 85748 Garching, Germany}

\author{K.\ Rolfs}%
\affiliation{Laboratory for Scientific Developments and Novel Materials, Paul Scherrer Institute, 5232 Villigen PSI, Switzerland}

\author{D.\ A.\ Tennant}
\affiliation{Oak Ridge National Laboratory, TN 37831 Oak Ridge, USA}

\author{K.\ Habicht}
\affiliation{Helmholtz-Zentrum Berlin f\"ur Materialien und Energie GmbH, 14109 Berlin, Germany}

\author{G.\ S.\ Uhrig}
\email{goetz.uhrig@tu-dortmund.de}
\affiliation{Lehrstuhl f\"{u}r Theoretische Physik I, Technische Universit\"at Dortmund, Otto-Hahn Stra\ss{}e 4, 44221 Dortmund, Germany}

\date{\rm\today}

\begin{abstract}
In this article we investigate the time dependence of the gap mode of \tcb{copper nitrate} at various temperatures. 
We combine state-of-the-art theoretical calculations with high precision \tcb{neutron resonance} spin-echo measurements to understand the anomalous decoherence effects found previously in this material. It is shown that the time domain offers a complementary view on this phenomenon, which allows us to directly compare experimental data and theoretical predictions without the need of further intensive data analysis, such as (de)convolution. 
\end{abstract}


\maketitle

\section{Introduction}

\tcr{Understanding finite temperature correlations in quantum magnets quantitatively is an ongoing, largely unsolved problem. Complex interactions in these materials lead to a variety of phases with manifold correlations, which are measurable at temperatures very low compared to the energy scale of the system.}
A plausible expectation is that quantum effects are purely suppressed upon increasing temperature due to the additional thermal fluctuations. \\
However there is a large set of counter-examples, where the interplay of quantum and thermal fluctuations lead to interesting new phenomena \cite{Sachdev:QPT} or that increasing temperature leads to a more robust phase \cite{Schmid2002}. \\
Besides static correlations at zero frequency, the finite frequency response is an interesting quantity, since it results from excitations in the system. 
In gapped quantum magnets, the energetically lowest excitations are coherent at zero temperature and do not decay. This is signaled by a $\delta$-function in frequency space. Hence, they can be viewed as infinitely longlived quasi-particles. At low temperature an exponentially small, but finite, number of these quasi-particles is present in the system, leading to collisions between them. These scattering processes transfer momentum and induce a loss of coherence for a single excitation.\\
A purely statistical model to describe the collisions leads to an exponential decay of the coherence, corresponding to a symmetric Lorentzian line-shape in frequency space. Indeed many studies on the spin dynamics in Heisenberg antiferromagnets suggested a universal Lorentzian line-shape for gapped as well as gapless excitations \cite{harris71,ty90,bayra06,Sachdev98,huber08,bayra13} at low temperature. \\
However, recent investigations of one dimensional \cite{tennant12a, klyushina2016} and three dimensional \cite{lake12a} materials of coupled spin dimers using inelastic neutron scattering (INS) show, that the line-shape in the frequency domain develops an asymmetric tail. It has been argued that such tails may be found in a broad range of quantum systems. The concomitant asymmetry, cannot be described by a pure Lorentzian at finite temperatures. Hence, the details of the scattering processes have a considerable impact on the correlations and more sophisticated theories have been employed to explain this phenomenon \cite{fabri97a,mikes06,essler08a,essler08b,essler09a,essler10a,lake13,jense14,tiege14,streib2015,klyushina2016,fause14}. \\
In the present paper, we explore a complementary path by analyzing the single quasi-particle correlations directly in the time domain, \tcr{the natural domain of decaying correlations}. \tcb{We compare the results to a statistical model, which predicts a Lorentzian line shape, where the temperature dependence enters only in the width of the peak \cite{Sachdev98}.}  \\
Using conventional INS it has been shown that copper nitrate (Cu(NO$_3$)$_2 \cdot 2.5$ D$_2$O), a model material for a one dimensional alternating Heisenberg chain (AHC), displays a temperature dependent asymmetric line-shape in the frequency domain \cite{xu2000, tennant03, tennant12a}. The material was recently investigated in the time domain using the high resolution neutron resonance spin-echo triple-axis \tcr{spectroscopy} (NRSE-TAS)  \cite{groitl2016}. The advantage of NRSE-TAS \cite{Keller2002, Habicht2004} is, that it gives access to slow processes corresponding to energies in the $\mu$eV range, which are not accessible by conventional INS. Contrary to conventional \tcr{triple-axis spectroscopy (TAS) and time-of-flight (ToF) INS}, where inaccuracies in background correction directly affect the linewidth and asymmetry, the linewidth or asymmetry measured with NRSE-TAS is insensitive to the background intensity, since the broad distribution in energy of the background is depolarized and does not contribute. In addition, for spin-echo no deconvolution of the data with the instrumental resolution is necessary; it reduces to a simple normalization of the raw data.\\
In a previous study of the material \cite{groitl2016}, the experimental results in the time domain were analyzed by fitting it to a phenomenological formula \cite{tennant12a}. This formula models an asymmetric line-shape in frequency domain and was used to compare NRSE-TAS and ToF data to demonstrate the capabilities of NRSE-TAS.\\
In the present paper we use the diagrammatic Br\"uckner approach \cite{fause14,fause15,klyushina2016} to capture thermal fluctuations in the system. This approach has been explicitly developed to calculate finite temperature effects on the single particle spectrum in gapped systems. \\
By combining the theoretical prediction of the Br\"uckner approach with the measurements of NRSE-TAS, we show, that the Br\"uckner approach is able to capture the dynamics of thermal fluctuations quantitatively without any additional fitting.
The theory reveals that the hard-core property of the excitations is the main source for the decoherence observed. This paves the way for experimental as well as theoretical studies to explore anomalous decoherence in quantum magnets directly in the time domain. \\
The article is set up as follows. In the next section, we briefly introduce the diagrammatic Br\"uckner approach to calculate single particle spectral functions at finite temperature. The NRSE-TAS technique is explained in section \ref{sec.experiment}. The comparison of theory and experiment takes place in section \ref{sec.comparison} \tcr{ and section \ref{sec.summary} concludes our article.}

%
%
%
%
%
%
%
%
%
%

\section{Theory: Diagrammatic Br\"uckner approach}

\begin{figure}[]
\centering
\includegraphics[width=0.99\columnwidth]{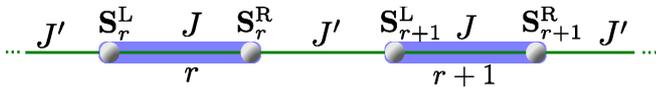}
\caption{Sketch of the couplings in the Hamiltonian of the alternating Heisenberg chain.}
\label{fig.AHC}
\end{figure}

\begin{figure}[]
\centering
\vspace*{12pt}
\includegraphics[width=0.9\columnwidth]{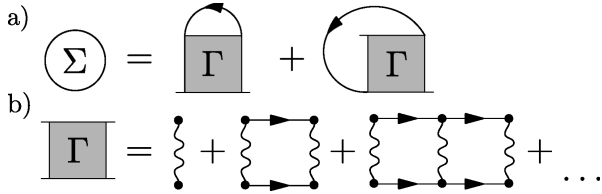}
\caption{a) Self-energy diagrams in leading order in $\exp(-\beta\Delta)$. The first diagram generates Hartree-like diagrams and the second the Fock-like diagrams. \\
b) Definition of the renormalized effective interaction $\Gamma$ between the excitations at finite temperature.}
\label{fig.self_energy}
\end{figure}

Copper nitrate is a model material for the spin-$1/2$ AHC \cite{xu2000}. Its Hamiltonian reads
\begin{align}
H = \sum\limits_{r} J \mathbf{S}_{r}^{\mathrm{L}} \cdot \mathbf{S}_{r}^{\mathrm{R}} + J' \mathbf{S}_{r}^{\mathrm{R}} \cdot \mathbf{S}_{r+1}^{\mathrm{L}}.
\end{align}
A graphical representation of the Hamiltonian is given in Fig.\ \ref{fig.AHC}. The index $r$ denotes dimers with strong interaction $J$. The dimers are coupled by a weaker interaction $J'$.\\
For copper nitrate the coupling constants are $J=0.443\,$meV and $J' = 0.101\,$meV  \cite{tennant12a}. Hence the alternation ratio $\alpha = J'/J \approx 0.227$ is rather small and the ground state of the chain is close to the product state with singlets on each dimer, $\prod_r \ket{s}_r = \prod_r \left( \ket{\uparrow \downarrow}_r - \ket{\downarrow \uparrow}_r \right)/\sqrt{2}$. The gap of the first excitations is given by $\Delta = 0.385\,$meV $\approx 4.5\,$K$k_\mathrm{B}$. \\
First, we introduce triplon operators to obtain a quasi-particle picture of the Hamiltonian \cite{Sachdev1990}.
The excitations of the AHC are threefold degenerate triplon states, which are created by the triplon operators $t_{r,\gamma}^\dagger, \, \gamma \in \lbrace x,y,z \rbrace \,$, where $\gamma$ denotes the flavor of the triplon. The operators fulfill a hard-core bosonic commutator relation, i.e., they commute on different dimers, but on a single dimer only one excitation is allowed at maximum. 
If the interaction $J'$ is switched on\tcb{,} quantum fluctuations are induced and the triplon excitations become dispersive. \\
To capture the zero temperature quantum fluctuations quantitatively we apply a continuous unitary transformation \cite{wegne94,knett00a,knett03a,kehre06,fisch10a} to map the Hamiltonian onto an effective Hamiltonian that conserves the number of triplons in the system. To first order in the parameter $\alpha$ the effective Hamiltonian is given by
\begin{equation}
\label{eq.interaction_in_effective_Hamiltonian} 
\begin{aligned}
&\frac{H_\mathrm{eff}}{J} = E_0 + \sum\limits_r \sum\limits_\gamma t_{r,\gamma}^\dagger t_{r,\gamma}^{\phantom\dagger}  
  - \frac{\alpha}{4} \sum\limits_r \sum\limits_\gamma t_{r,\gamma}^\dagger t_{r+1,\gamma}^{\phantom\dagger}  + \mathrm{h.c.}\\
  &+ \frac{\alpha}{4} \sum\limits_r \sum\limits_{\gamma \neq \phi} \left( t_{r,\gamma}^\dagger t_{r,\phi}^{\phantom\dagger} t_{r+1,\phi}^\dagger t_{r+1,\gamma}^{\phantom\dagger} - t_{r,\gamma}^\dagger t_{r,\phi}^{\phantom\dagger} t_{r+1,\gamma}^\dagger t_{r+1,\phi}^{\phantom\dagger} \right) 
\end{aligned}
\end{equation}
In our calculations we used the effective Hamiltonian up to order $6$, which is sufficient to capture all quantum effects in the model. \\
On top of this effective Hamiltonian we apply the Br\"uckner approach to calculate the finite temperature effects on the spectrum. For quantum fluctuations in quantum magnets the approach was first introduced in {Ref.\ }\cite{kotov98a}. It was extended in {Ref.\ }\cite{fause14} to describe the effects of thermal fluctuations and continued to multi-flavor systems, such as triplon models, in {Ref.\ }\cite{fause15}. \\
The systematic control parameter of the theory is the low density of thermally excited hard-core bosons
which is proportional to $\exp(-\beta \Delta)$ in a gapped system where $\beta$ is the inverse temperature.
Since there exists no perturbation theory for hard-core bosons, the excitations are treated as normal bosons, but subject to an infinite on-site interaction to inhibit double occupancies. \\
The quantity of interest for inelastic neutron scattering is the dynamic structure factor which is related to the imaginary part of the Green function by means of the fluctuation-dissipation theorem
\begin{align}
\label{eq.fluctuation_dissipation}
S^{T>0}_{zz}(p, \omega) = \frac{1}{1-e^{-\beta\omega}} \frac{1}{\pi} \mathrm{Im} \left[ G^{zz}(p, \omega) + G^{zz}(p, -\omega) \right],
\end{align}
where $p$ is the momentum in chain direction.\\
The single quasi-particle peak is the dominant contribution to the dynamic structure factor \cite{schmi03c} and corresponds to the observable measured in NRSE-TAS. Hence we restrict the calculations to the single particle Green function $G^{zz} \propto \left\langle \left[ t_{p,z}^{\phantom\dagger}, t_{p,z}^\dagger \right]\right\rangle$ and leave out the temperature dependence of modes with higher quasi-particle number and modes at zero frequency. \\
Vertex corrections \tcb{treated in Ref.\ \cite{exius10}} are left out because their effects are negligible small in the parameter regime investigated.\\ 
To obtain the imaginary part of the Green function, we first calculate the single particle self energy. The first order contribution in the parameter $\exp(-\beta \Delta)$ is given by the diagrams in Fig.\ \ref{fig.self_energy} a), translating to
\begin{subequations}
\begin{align}
\label{eq.self_energy1}
\Sigma^{\gamma \gamma}(P) = \frac{1}{N} \sum\limits_\phi \sum\limits_{K} (1+\delta_{\gamma,\phi}) G_0^{\phi \phi}(K) \Gamma^{\gamma,\phi}(P+K), \\
\label{eq.bethe_salpeter}
\Gamma^{\gamma,\phi}(P) = \lim\limits_{U \rightarrow \infty} \frac{U}
{N \beta + U \sum\limits_K G_0^{\gamma \gamma}(P+K) G_0^{\phi \phi}(-K)},
\end{align}
\end{subequations}
where $\Sigma^{\gamma \gamma}(P)$ is the frequency-dependent self-energy, $N$ denotes the total number of sites, $P$ and $K$ are 2-momenta, i.e.,\ $P = \left( p, i \omega_p \right)$, and $G^{\phi \phi}_0$ is the bare Green function.
We obtain the effective interaction $\Gamma$ between the hard-core bosons at finite temperature by summing all diagrams given in Fig.\ \ref{fig.self_energy} b). \\
Note that the interaction vertices represent the local repulsion $U$, which is sent to infinity analytically in the end. 
The sum of all ladder interactions in Fig.\ \ref{fig.self_energy} b) can be calculated as a geometric series and is called the Bethe-Salpeter equation \eqref{eq.bethe_salpeter}.
The effective interaction includes two contributions: one from the low energy sector of two separate quasi-particles, and one from an anti-bound state of two quasi-particles with energy $\omega \propto U$. Although $U$ is sent to infinity, the anti-bound state has a subtle effect on the low energy physics which must be taken into account \cite{fause14,fause15}.
\\
To also take the additional interactions in the effective Hamiltonian into account, we apply a self-consistent Hartree-Fock decoupling \cite{streib2015}.
The decoupling has no effect on the imaginary part of the self energy, but it shifts the peak positions slightly. Thus it has no significant effect on the line-shape of the excitations and the hard-core interaction is the primary source for the broadening of the single quasi-particle peak.
Once the self-energy is computed, we can directly obtain the frequency-dependent structure factor from Eq.\ \eqref{eq.fluctuation_dissipation}.
\begin{figure}[]
\centering
\includegraphics[width=0.99\columnwidth]{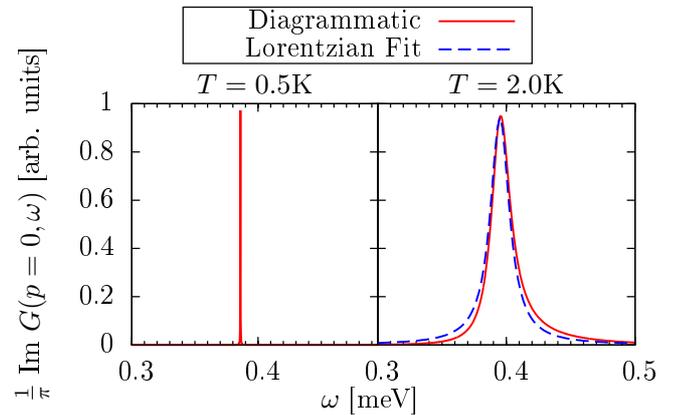}
\caption{Single particle spectral function Im $G(p=0,\omega)$ for the AHC as function of $\omega$ for $T=0.5$K and $T=2$K. \tcb{The full red line is the result from the diagrammatic Br\"uckner approach, while the blue dashed line shows a comparison to a pure Lorentzian line shape.} The y-axis has been scaled to make the plots comparable.}
\label{fig.Lineshapes}
\end{figure}\\
Fig.\ \ref{fig.Lineshapes} shows the spectral function, as obtained from the Br\"uckner approach for $T=0.5$K and $T=2$K \tcb{in comparison with a pure Lorentzian line shape, using a decay rate determined in section \ref{sec.comparison}.} For very low temperatures the response is sharply peaked at the position of the dispersion. As temperatures increases, the width of the peak increases, signalling a decrease in coherence. \tcb{Note that the signal obtains an asymmetric tail towards higher energies at $T=2$K. This phenomenon was already reported in Ref.\ \cite{tennant12a} for copper nitrate and in Ref.\ \cite{klyushina2016} for BaCu$_2$V$_2$O$_8$ using conventional INS.}\\
Since the NRSE-TAS method captures the correlations in time domain at fixed momentum $p$ and as function of spin-echo time $\tau$ (see next section), we Fourier transform the theoretical data to the time domain, yielding the intermediate scattering function $I(p,\tau)$. Then, its envelop is compared to the experimental data. \\
Due to the high resolution of the theoretical spectral function, the numerical Fourier transformation provides highly accurate data for the time range measured in experiment.


\section{Experiment: Neutron spin-echo spectroscopy}
\label{sec.experiment}
For the sake of brevity only a brief introduction will be given to the method itself and we refer for the detailed description to the literature \cite{Mezei21980,HabichtNSEbook2003_mod,Golub1993,Gaehler1998,Felber1998,Habicht2003,groitl2016}. Neutron spin-echo measures the polarization of the neutron beam $P=\left\langle \sigma_x\right\rangle$, which is called the echo amplitude. The two spin states are associated with two correlation volumes \cite{HabichtNSEbook2003_mod,Felber1998} and a precession region before the sample (see Fig. 1 in Ref. \cite{groitl2016}) is used to split the two states \cite{Golub1993} by a relative time delay $\tau$, the so-called spin-echo time. The spin states scatter at the sample at times $t$ and $t+\tau$, where $\tau$ is identical to the van-Hove correlation time \cite{Gaehler1998}. In a second, reversed precession region after the sample the time delay is canceled out and the states interfere at the detector. Thus, the echo amplitude $P$ is a direct measure of the time dependence of the intermediate scattering function $I(\mathbf{Q},\tau)$.\\
For quasielastic experiments classical neutron spin-echo (NSE) instruments based on DC fields were very successful. It has been shown, that for the investigation of dispersive excitations such as phonons and magnons it is necessary to combine spin-echo and \tcr{ TAS techniques } to select small regions in $(\mathbf{Q},\omega)$ space. The experiments described in Ref. \cite{groitl2016} have been carried out at the NRSE-TAS spectrometer TRISP \cite{Keller2002} at the FRM II. \\
A high quality crystal of Cu(NO$_3$)$_2 \cdot 2.5$ D$_2$O with a sample mass of $4\,$g and a deuteration ratio of $>99.38\,$\% was used. The \tcr{one-triplon} mode was studied in the minimum of the dispersion at \tcr{ $\mathbf{Q}=(1 0 1)\,$ } r.l.u., $\hbar\omega_0=0.385\,$meV. This corresponds to the $p=0$ mode in the pure 1D theoretical picture. By varying the length of the second precession region a sinusoidal variation of the count rate corresponding to a $2\pi$ rotation of the neutron spins is recorded. The echo amplitude and the intensity are then extracted by using Eq. (3) in Ref. \cite{groitl2016}. The echo amplitude was measured for different spin-echo times and \tcr{for four temperatures} with the data shown in Fig. \ref{fig.Data_Comparison}. The data was corrected for background and instrument resolution. Note that the rather large \tcr{error bars} are purely statistical and \tcr{do not} contain any systematic error arising from a deconvolution with the resolution function. Due to the equal distribution between non-spin-flip (NSF) and spin-flip (SF) scattering for copper nitrate $P(\tau=0)=0.5$ is expected, which is in very good agreement with the recorded data \cite{groitl2016}.

\section{Comparison}
\label{sec.comparison}
\begin{figure}[h!]
\centering
\includegraphics[width=0.99\columnwidth]{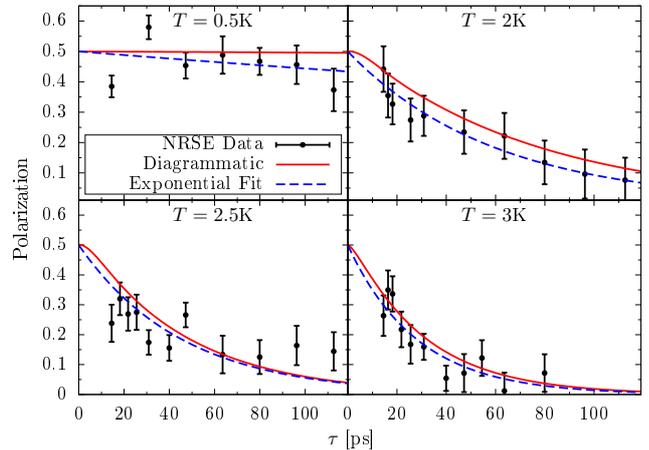}
\caption{Echo amplitude as function of the spin-echo time $\tau$ for four different temperatures $T$. The dots are the experimental data points obtained from the NRSE-TAS measurements, \tcb{full red} lines indicate the theory curve from the Br\"uckner-approach. \tcb{ The blue dashed lines are exponential fits of the form $P(\tau) = 0.5 \cdot \exp(-\gamma \tau)$. The fit parameters for the different temperatures are $T=0.5 \mathrm{K} \,: \, \gamma = 0.00119 \mathrm{ps}^{-1}$, $T=2.0 \mathrm{K} \,: \, \gamma = 0.01681 \mathrm{ps}^{-1}$, $T=2.5 \mathrm{K} \,: \, \gamma  = 0.02193 \mathrm{ps}^{-1}$ and $T=3.0 \mathrm{K} \,: \, \gamma = 0.03559 \mathrm{ps}^{-1}$.}}
\label{fig.Data_Comparison}
\end{figure}
Fig.\ \ref{fig.Data_Comparison} shows a comparison of the experimental data points for the echo amplitude $P(\tau)$ with the theoretical results \tcb{of the Br\"uckner approach and an exponential fit, as expected from a pure Lorentzian line shape in frequency space in the statistical model.} Four different temperature sets were measured to investigate the temperature dependence of the spin-echo signal and thus the temporal correlations of the excitation. Taking the equal distribution of NSF \tcr{and} SF scattering into account the theoretical curves were normalized to $P(\tau = 0) = 0.5$.\\
For $T=0.5K$, close to the zero temperature limit, the excitation is still long-lived and shows no decay in coherence up to $112\,$ps corresponding to a minimal energy resolution of $\hbar/(112\, \mathrm{ps}) \approx 5 \, \mu\mathrm{eV}$. As temperature increases, the echo amplitude shows a faster decay due to additional scattering processes and the experimental data points show a good agreement with the theoretical results up to $T=3K\approx0.67\Delta/k_\mathrm{b}$. We stress that the Br\"uckner theory has no free parameters and matches to the experiment without any fitting at all.\\
 \tcb{The pure exponential fits, however, are based on four fit parameters $\gamma$, one for each temperature. They agree slightly better than the parameter free Br\"uckner theory. Hence the process responsible for the decay can not be determined purely from the polarization.} \\
\tcb{In the Br\"uckner theory} the deeper reason for the decay lies in the imaginary part of the self energy. Since the self energy is determined by the effective interaction $\Gamma$, non-trivial scattering processes lead to the specific form of the decay. A similar effect is observed in Ref.\ \cite{fause14} and is a consequence of the hard-core bosonic nature of the excitations.\\ Note, that the theoretical prediction shows a vanishing slope for $P(\tau = 0)$ \tcb{in contrast to the pure exponential decay}. This is expected in a full quantum mechanical description, where the decay channel has only a finite support in frequency. Due to the finite minimal correlation time $\tau_{\mathrm{min}} \approx 14.5$ps, which is measurable, this feature is hardly examinable in the experiment. \\
\begin{figure}[]
\centering
\includegraphics[width=0.99\columnwidth]{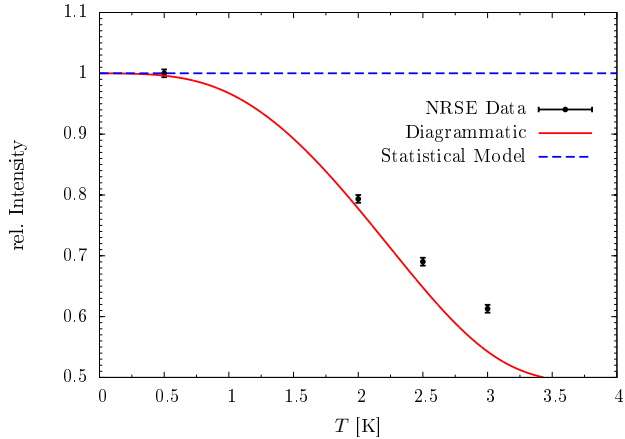}
\caption{\tcb{Comparison between the experimentally obtained intensity (black data points) with the prediction from the diagrammatic Br\"uckner approach (full red line) and the constant intensity in the statistical model (blue dashed line).} The experimental data was normalized to the base temperature value and is in agreement with the \tcb{Br\"uckner} theory except that the experiment shows a slightly slower decay in intensity.}
\label{fig.Intensity_Comparison}
\end{figure}
\tcb{ Since the polarization, is not sufficient to distinguish between a simple statistical model and the results of the Br\"uckner theory, we pass to another quantity which is accessible to theory and experiment: the intensity of the mode. The intensity of the mode is directly connected to the quasi-particle residue. In a statistical description of the decay, the quasi-particle residue is independent of temperature and stays constant \cite{Sachdev:QPT, Sachdev98}. On the other hand the Br\"uckner theory predicts a strong dependence of the total intensity on temperature. The more hard-core bosons are thermally excited in the system, the more it is difficult to add further bosons by neutron scattering. Thus the residue decreases significantly upon increasing temperature, see Eq.\ (9) in Ref.\ \cite{fause15}. }
Figure \ref{fig.Intensity_Comparison} shows a comparison between the intensity measured by NRSE-TAS and the theoretical prediction \tcb{of the Br\"uckner theory}. The experimental data is normalized to the low temperature regime at $T=0.5$K. \\
The agreement between the Br\"uckner theory and experiment is remarkable. The slight deviation at $T=3$K is in agreement with the low-temperature expansion applied in the Br\"uckner approach. \tcb{The pure statistical model fails to explain the temperature dependence of the intensity.}
As before, the prediction \tcb{of the Br\"uckner theory} results from the full quantum model in Eq.\ \eqref{fig.AHC} and has no free parameters.
The decreasing intensity results from the hard-core bosonic commutator relation and is not related to the temperature dependence in the fluctuation-dissipation theorem in Eq.\ \eqref{eq.fluctuation_dissipation}, see also Refs.\ \cite{fause14, fause15}. \tcb{Thus we can identify the hard-core bosonic scattering processes as the main source for the decay.}
\tcr{\section{Summary}
\label{sec.summary}
}
In summary, we investigated the temperature dependence of the gap mode of copper nitrate, a realization of \tcr{an} alternating Heisenberg chain, in the time domain. On the experimental side, we used NRSE-TAS measurements to obtain high resolution data of the time-dependent spin-echo signal. On the theoretical side, we used the diagrammatic Br\"uckner approach to calculate the single particle spectral function of the AHC, which was transformed into the time domain to model the time-dependence of the echo-amplitude. Very good agreement between theory and experiment has been achieved without any extensive data analysis or \tcr{variation of parameters.}\\
On basis of the polarization and intensity of the mode \tcb{ it was argued, that the single particle decoherence cannot be described by a pure statistical model}, but rather by non-trivial scattering processes of the hard-core bosonic excitations. Hence quantum corrections, which are induced by the hard-core property, play an important role in describing the physics of quantum magnets at finite temperature.\\
Our work shows, that the direct analysis in the time domain is a promising route to understand dynamic correlation and to deepen our understanding of quantum coherence at finite temperature. An interesting question is, whether it is possible to increase the fraction of non-trivial scattering processes to further enhance the quantum character of the excitations at finite temperature. 

%
%


\begin{acknowledgments}
We acknowledge financial support of the Helmholtz Virtual Institute ``New states of matter and their excitations''. B.F.\ acknowledges the Fakult\"at Physik of the TU Dortmund University for funding in the context of the ``Bestenf\"orderung''. 
\end{acknowledgments}


\bibliographystyle{apsrev}

\end{document}